\documentclass[12pt, a4paper]{article} 
\usepackage[utf8]{inputenc}
\usepackage{authblk}
\usepackage[nodisplayskipstretch]{setspace}
\usepackage[margin=1.0in]{geometry}
\usepackage{graphicx}
\graphicspath{ {./figures/} }
\usepackage{subcaption}
\usepackage{amsmath, amssymb, amsthm}
 \usepackage{mathrsfs}
\usepackage{lineno}

\usepackage{sectsty}
\sectionfont{\large}

\usepackage{float}
\usepackage[figuresright]{rotating}

\usepackage{booktabs}
\usepackage{hyperref}
\usepackage{bm}

\usepackage{xcolor}

\usepackage{algorithm}
\usepackage[noend]{algpseudocode}

\newtheorem{theorem}{Theorem}
\usepackage{enumitem}
\newlist{enumsteps}{enumerate}{2}
\setlist[enumsteps,1]{label=\textbf{Step \arabic*}}

\providecommand{\keywords}[1]
{
  \small	
  \textbf{\textit{Keywords: }} #1
}

\usepackage{natbib}

\title{\textbf{\large  Estimation of Heterogeneous and Functional Cox Model}}

\author[1]{\small Yichen Lou}
\author[2]{\small Mingyue Du}
\author[3]{\small Yuxiang Wu}
\author[4]{Jianguo Sun}

\affil[1]{School of Physical and Mathematical Sciences, Nanyang Technological University, Singapore.}
\affil[2]{School of Mathematics, Jilin University, Changchun, China.}
\affil[3]{Department of Biostatistics, Emory University, Atlanta, U.S.A.}
\affil[4]{Department of Statistics and Data Science, Southern University of Science and Technology, Shenzhen, Guangdong, China.}
\affil[*]{Address correspondence to: suncolumbia@163.com} 

\date{}

\date{}

\begin{document}

\maketitle


\begin{abstract}
Cox or proportional hazards model is the most commonly used regression model for failure time data and has been extended in various ways.
In this paper, we consider inference for a heterogeneous and functional Cox model with the focus on estimation of the latent heterogeneous or subgroup structure
based on interval-censored failure time data.  For the problem, we propose a novel sieve maximum likelihood estimation procedure
with the uses of B-splines and Bernstein polynomials. 
For the implementation of the proposed method, an EM algorithm is developed and the number of subgroups can be determined in a data-driven manner. 
Also the resulting estimators are shown to be consistent and asymptotically normal, and an extensive simulation study shows that the proposed method works well in practice.  Finally, we apply it to an Alzheimer's Disease 
study that motivated this investigation and provide some new insights about the roles of imaging and clinical factors in 
the progression of Alzheimer's disease.
\end{abstract}



\doublespacing
\setstretch{2.0}

\section{Introduction}

Cox or proportional hazards model is the most commonly used regression model for failure time data.  
On the other hand, it is well-known that it has some limitations and thus many extensions have been proposed.
In this paper, we consider a heterogeneous and functional Cox model, a situation that has not been investigated, 
with the focus on estimation of the latent heterogenous or subgroup structure based on interval-censored failure time data.
Among others, one situation where such models play essential roles is precision medicine. 
Also it has been shown that ignoring the heterogeneity may lead to biased estimation and misleading conclusions \citep{pei2024latent}.

By interval-censored failure time data, we mean that the failure time of interest cannot be exactly observed and instead 
is known only to belong to an interval \citep{sun2006statistical, sun2022emerging}. Such data frequently occur in,
for example, periodical follow-up studies such as clinical trials and epidemiological studies. 
It is well-known that right-censored data, the type of failure time data studied most in the literature, can be seen 
as a special case of interval-censored data and the analysis of the latter is much more challenging than the former due to
the more complicated structures of the latter.  For example, under the proportional hazards model, 
the simple partial likelihood commonly used for the former is not available anymore for the latter. 

An example of interval-censored data that motivated this study is given by the Alzheimer's Disease Neuroimaging Initiative (ADNI), 
a large longitudinal and multicenter study initiated in 2004 \citep{weiner2013alzheimer}.   For the study, one of the primary objectives is 
to identify biomarkers for the early detection and monitoring of Alzheimer's disease (AD), and among others, one variable of interest is 
the time from mild cognitive impairment (MCI) to AD conversion 
 \citep{weiner2015impact}.  Since the participants were assessed only at discrete follow-up visits, the exact conversion time 
 cannot be observed and only interval-censored data on it are available \citep{risacher2009baseline}. 
 In addition, substantial heterogeneity may exist among individuals with MCI because of the differences in disease progression and 
 underlying biological characteristics \citep{apostolova2006conversion}.   Furthermore, ADNI provides a rich collection of clinical, imaging, 
 genetic, and biochemical measurements plus some functional covariates such as hippocampal shape, which 
 has been shown to be closely associated with AD progression \citep{kong2018flcrm}. More details on the study will be given below. 

Functional covariates, which often arise from longitudinal measurements and neuroimaging data, have received increasing attention in survival analysis. Due to their infinite-dimensional nature, some special statistical methods are usually needed for their analysis. Existing approaches include Bayesian methods \citep{lee2015bflcrm} and kernel-based methods \citep{hao2021semiparametric,qu2016optimal}. Functional principal component analysis has also been widely used for dimension reduction by representing functional covariates through the Karhunen-Lo\'eve expansion and reducing functional Cox models to finite-dimensional models \citep{chen2025high,kong2018flcrm}. As an alternative, B-spline-based approaches have been considered \citep{gellar2015cox}, and among others,  \cite{liu2025efficient} recently studied such an approach for the accelerated failure time model with right-censored data. One advantage of the spline-based approach is that it directly approximates the functional coefficient and avoids truncating the functional covariate to only a small number of leading principal components. 
However, these methods do not consider the possible subgroup structure together with interval-censored failure time data.

As mentioned above, a large literature has been established for subgroup analysis. In general, the existing methods can be classified into two broad classes. One is the fusion penalty-based approach. For example, \citet{ma2017concave} proposed a pairwise fusion penalization method for linear regression that can identify the subgroup structure without requiring the number of subgroups to be specified in advance. This idea has subsequently been extended to Poisson regression \citep{chen2019subgroup}, quantile regression \citep{zhang2019robust,lu2021multiply}, additive partially linear models \citep{liu2019subgroup,cai2024subgroup}, panel data models \citep{wang2024homogeneity}, and multivariate response regression \citep{wu2025simultaneous}. The other is the mixture model-based approach, including Gaussian mixture models \citep{wei2013latent}, low-rank methods for mixtures of linear regressions \citep{chaganty2013spectral}, and logistic-normal mixture models \citep{shen2015inference}. Compared with the fusion penalty-based methods, the mixture model-based methods are often relatively straightforward to implement and naturally provide probabilistic subgroup membership.

For failure time data, however, relatively less work exists on subgroup analysis. Heterogeneous proportional hazards and accelerated failure time models have been developed to identify latent subgroup structures in survival data \citep{h2021subgroup,y2021subgroup}, and some extensions have been developed for high-dimensional settings \citep{h2022histopathological}, functional covariates \citep{z2022subgroup}, and recurrent event outcomes \citep{z2022semiparametric}. However, most of these methods focus on right-censored data, and as mentioned above, interval-censored data have a much more complicated likelihood structure, and the usual partial likelihood is not available. 
Although several authors, including \cite{huang2022subgroup} and \cite{du2026subgroup}, have discussed subgroup analysis for interval-censored data,
none has considered functional covariates. 
In other words, there does not seem to exist an established procedure that can simultaneously accommodate interval censoring, functional covariates, and latent subgroup heterogeneity.

To address these issues, we propose a functional subgroup analysis procedure for interval-censored data under a heterogeneous proportional hazards model. For estimation,
the sieve maximum likelihood estimation is employed with the uses of B-spline basis functions and Bernstein polynomials to approximate the subgroup-specific 
functional coefficients and the unknown baseline cumulative hazard functions, respectively. For the implementation, latent variables are introduced to represent subgroup membership and an EM algorithm is developed with the number of subgroups selected in a data-driven manner. Also, the asymptotic properties of the resulting estimators are established, including consistency and asymptotic normality.

The remainder of this paper is organized as follows. In Section~2, after introducing the observed data and notation, we first present 
the proposed heterogeneous proportional hazards model with functional covariates and the resulting likelihood function.
Then the proposed sieve maximum likelihood estimation and EM algorithm will be described.  The asymptotic properties of the resulting estimators
are established in Section~3, and Section~4 presents some results obtained from a simulation study conducted to assess the finite sample performance 
of the proposed method. In Section~5, we apply the proposed method to the ADNI 
described above, and Section~6 concludes with some discussion and remarks.

\section{Sieve Maximum Likelihood Estimation}

\subsection{Heterogeneous and Functional Cox Model and Likelihood Function}

Consider a failure time study involving $n$ independent subjects and let $T_i$ denote the failure time of interest for subject $i$. 
Suppose that the subjects belong to $K$ potentially heterogeneous latent subgroups, and let $G_i\in\{1,\ldots,K\}$ denote the latent
subgroup membership of subject $i$. For each subject, let ${\bm W}_i\in\mathbb R^{p_1}$ and ${\bm Z}_i\in\mathbb R^{p_2}$ 
denote two vectors of covariates and $X_i(\cdot)$ a functional covariate.  Here ${\bm W}_i$ contains the covariates 
whose effects on $T_i$ are common across subgroups, whereas ${\bm Z}_i$ contains the covariates whose effects may vary across subgroups. 
In the following, we assume that given all of the covariates and $G_i=k$, $T_i$ follows the heterogeneous and functional proportional hazards model
\begin{equation}\label{haz}
\Lambda_i\{t\mid G_i=k,{\bm W}_i,{\bm Z}_i,X_i(\cdot)\} =
\Lambda_{k}(t)\exp\left\{{\bm W}_i^\top{\bm \beta}+{\bm Z}_i^\top{\bm \alpha}_k+\int_{\mathcal I_0}X_i(s)\gamma_k(s)\,ds\right\}\, 
\end{equation}
in terms of the cumulative hazard function. 
In the above, $\Lambda_{k}(t)$ denotes an unknown subgroup-specific baseline cumulative hazard function, ${\bm \alpha}_k\in\mathbb R^{p_2}$ 
the subgroup-specific effect of ${\bm Z}_i$, $\gamma_k(\cdot)$ the subgroup-specific functional coefficient, and ${\bm \beta}\in\mathbb R^{p_1}$ 
the common effect of ${\bm W}_i$. 

Define ${\bm \alpha}=({\bm \alpha}_1^\top,\ldots,{\bm \alpha}_K^\top)^\top$,  ${\bm \gamma}(\cdot)=\{\gamma_1(\cdot),\ldots,\gamma_K(\cdot)\}^\top$ and  ${\bm \Lambda}(\cdot)=\{\Lambda_{1}(\cdot),\ldots,\Lambda_{K}(\cdot)\}^\top$. 
Without loss of generality, we assume that the functional covariate has been centered and that $\mathcal I_0=[0,1]$. 
It is worth noting that model~\eqref{haz} includes some commonly used models as special cases. If ${\bm\alpha}_k$ and $\gamma_k(\cdot)$ 
are assumed to be common across subgroups, for example, the model reduces to the standard Cox model with a functional 
covariate \citep{wen2014functional, shi2022functional}. If the functional covariate $X(s)$ is further omitted, the model reduces to 
the standard Cox model.

Suppose that the observed data have the form $\{(L_i,R_i],{\bm W}_i,{\bm Z}_i,X_i(\cdot)\}_{i=1}^n$, where $L_i<T_i\leq R_i$. 
That is, only interval-censored observations on the $T_i$'s are available. 
To characterize the latent subgroup membership, let $\omega_k=\mathbb P(G_i=k)$ denote the probability that a subject belongs to 
subgroup $k$ and define ${\bm \omega}=(\omega_1,\ldots,\omega_K)^\top$, ${\bm \vartheta} = ({\bm \beta},{\bm \alpha},{\bm \gamma},{\bm \Lambda})$ 
and ${\bm \theta}=({\bm \vartheta}, {\bm \omega})$.  Then we have that $0<\omega_k<1$ and $\sum_{k=1}^K\omega_k=1$.
In the following, we assume independent interval censoring given the covariates and latent subgroup membership.
Then the observed likelihood function has the form
\begin{equation}\label{like1}
\mathcal L({\bm \theta}) = \prod_{i=1}^n\left\{\sum_{k=1}^K\omega_k\mathcal L_{ik}({\bm \vartheta})\right\}
= \prod_{i=1}^n\left[\sum_{k=1}^K\omega_k \Big\{S_{ik}(L_i)-S_{ik}(R_i) \Big\}\right]\, ,
\end{equation}
where
$$
S_{ik}(t) = \exp\left[-\Lambda_{k}(t)\exp\left\{{\bm W}_i^\top{\bm \beta}+{\bm Z}_i^\top{\bm \alpha}_k+\int_0^1X_i(s)\gamma_k(s)\,ds\right\}\right]
$$
denotes the conditional survival function of $T_i$ given $G_i=k$ and the covariates.

\subsection{Sieve Estimation}

For the estimation of model \eqref{haz} or ${\bm \theta}$, it is apparent that a natural approach is to maximize the log likelihood function 
$\ell({\bm \theta})=\log\mathcal L({\bm \theta})$. However, it is easy to see that the direct maximization is difficult because it involves 
two unknown functions ${\bm \Lambda}(\cdot)$ and ${\bm \gamma}(\cdot)$. To address this issue, we propose a sieve estimation approach 
in which Bernstein polynomials and B-spline functions are used to approximate them, respectively.

More specifically, define the parameter space $\Theta=\mathcal B\otimes\Pi\otimes\left\{\bigotimes_{k=1}^K\Gamma_k\right\}\otimes\left\{\bigotimes_{k=1}^K\mathcal M_k\right\}$, where $\mathcal B=\left\{{\bm \wp}=({\bm \beta}^{\top},\, {\bm \alpha}^{\top})^{\top}\in\mathbb R^{p_1+Kp_2}:\|{\bm \wp}\|\leq M\right\}$
for some constant $M$, $\Pi=\left\{{\bm \omega}:\omega_k \in (0, 1),\ \sum \omega_k=1\right\}$, 
and $\Gamma_k=\left\{\gamma_k\in L_2[0,1]:\|\gamma_k\|_{L_2}\leq C_\gamma\right\}$
for some constant $C_\gamma$. In addition, $\mathcal M_k$ denotes the set of all nonnegative, continuous, and nondecreasing functions on $[u,\tau]$, 
where $[u,\tau]$ covers the observed failure times.   Corresponding to the function ${\bm \gamma}(\cdot)$,
define the sieve space
$$
\Gamma_{kn}=\left\{\gamma_{kn}(s)=\sum_{l=1}^{q_n}\phi_{lk}b_l(s)={\bm \phi}_k^{\top}{\bm b}(s):\|{\bm \phi}_k\|_2\leq C_\gamma M_{q_n}^{1/2}\right\}\, ,
$$
where ${\bm b}(s)=\{b_1(s),\ldots,b_{q_n}(s)\}^{\top}$ denotes a $q_n$-dimensional cubic B-spline basis on $\mathcal I_0$
and ${\bm \phi}_k=(\phi_{1k},\ldots,\phi_{q_nk})^{\top}$ the corresponding coefficient vector. 
Corresponding to the baseline cumulative hazard function ${\bm \Lambda}(\cdot)$, define the sieve space
$$
\mathcal M_{kn}=\left\{\Lambda_{kn}(t)=\sum_{j=0}^m\psi_{kj}^{*}B_j(t;m,u,v):\sum_{j=0}^m|\psi_{kj}^{*}|\leq m,\ 0\leq\psi_{k0}^{*}\leq\cdots\leq\psi_{km}^{*}\right\}\, ,
$$
where
$$
B_j(t;m,u,\tau)=\binom{m}{j}\left(\frac{t-u}{\tau-u}\right)^j\left(1-\frac{t-u}{\tau-u}\right)^{m-j},\qquad j=0,\ldots,m\, ,
$$
are the Bernstein basis polynomials of degree $m$ with $m\asymp n^\nu$ for some $\nu\in(0,1)$. 
Note that based on the above approximation, we can write the functional covariate component as
$\int_0^1X_i(s)\gamma_{kn}(s)\,ds={\bm \phi}_k^{\top}{\bm D}_i$, 
where ${\bm D}_i=(D_{i1},\ldots,D_{iq_n})^{\top}$ and
$D_{il}=\int_0^1X_i(s)b_l(s)\,ds$ for $l=1,\ldots,q_n$. 

Let ${\bm \zeta}=({\bm \zeta}_1^\top,\ldots,{\bm \zeta}_K^\top)^\top$ with ${\bm \zeta}_k=({\bm \alpha}_k^\top,{\bm \phi}_k^\top)^\top$
and define ${\bm V}_i=({\bm Z}_i^\top,{\bm D}_i^\top)^\top$.
Then the likelihood function given in \eqref{like1} can be approximated by
\begin{equation}\label{liken}
\mathcal L_n({\bm \theta})
= \prod_{i=1}^n\left\{\sum_{k=1}^K\omega_k\mathcal L_{ik,n}({\bm \vartheta})\right\}
= \prod_{i=1}^n\left[\sum_{k=1}^K\omega_k \Big\{S_{ik,n}(L_i)-S_{ik,n}(R_i)\Big\} \right]\, ,
\end{equation}
where
$$
S_{ik,n}(t)=\exp\left[-\Lambda_{kn}(t)\exp\left\{{\bm W}_i^{\top}{\bm \beta}+{\bm V}_i^{\top}{\bm \zeta}_k\right\}\right]\, .
$$
We propose to estimate ${\bm \theta}$ by its value that maximizes the log likelihood function $\ell_n({\bm \theta})=\log\mathcal L_n({\bm \theta})$ 
over the sieve space $\Theta_n=\mathcal B\otimes\Pi\otimes\left\{\bigotimes_{k=1}^K\Gamma_{kn}\right\}\otimes\left\{\bigotimes_{k=1}^K\mathcal M_{kn}\right\}$. 
Note that the monotonicity constraints on the Bernstein coefficients can be removed through the reparameterization
$\psi_{k0}^{*}=\exp(\psi_{k0})$ and $\psi_{kl}^{*}=\sum_{j=0}^l\exp(\psi_{kj})$ for $l \ge 1$.  
Also note that although the cubic B-spline basis functions are used above for simplicity, the B-splines of other degrees can also be 
employed and it is straightforward to adapt the proposed method accordingly.

\subsection{EM Algorithm}

Let $\widehat{\bm \theta}$ denote the estimator of ${\bm \theta}$ defined above.  For its determination or the maximization of 
$\ell_n({\bm \theta})$, we 
first assume that the number of the latent subgroups $K$ is known and develop an EM algorithm.
Define the latent subgroup indicator $\chi_{ik}= \mathbb{I}(G_i=k)$, which equals one if subject $i$ belongs to subgroup $k$ and zero otherwise. If the subgroup memberships were observed, the complete data would be
$O^\sharp=\{(L_i,R_i],{\bm W}_i,{\bm Z}_i,X_i(\cdot),{\bm \chi}_i\}_{i=1}^n$, 
where ${\bm \chi}_i=(\chi_{i1},\ldots,\chi_{iK})^\top$. The corresponding complete-data log likelihood function would have the form
$$
\begin{aligned}
\ell_n^\sharp({\bm \theta}) =\sum_{i=1}^n\sum_{k=1}^K\chi_{ik}\log\{\omega_k\mathcal L_{ik,n}({\bm \vartheta})\} =\sum_{i=1}^n\sum_{k=1}^K\chi_{ik}\log\omega_k +\sum_{i=1}^n\sum_{k=1}^K\chi_{ik}\log\mathcal L_{ik,n}({\bm \vartheta})\, .
\end{aligned}
$$

In the E-step of the EM algorithm, given the current estimate $\widehat{\bm \theta}^{(r)}$ and the observed data ${\bm O}=\{O_i\}_{i=1}^n$, 
we compute the conditional expectation of the complete-data log likelihood function 
\begin{equation}\label{Qfunc}
\begin{aligned}
\mathcal Q\left({\bm \theta}\mid{\bm O},\widehat{\bm \theta}^{(r)}\right)
&=\mathbb E\left\{\ell_n^\sharp({\bm \theta})\mid{\bm O},\widehat{\bm \theta}^{(r)}\right\} =\sum_{i=1}^n\sum_{k=1}^K\widetilde{\chi}_{ik}^{(r)}\log\omega_k
+\sum_{i=1}^n\sum_{k=1}^K\widetilde{\chi}_{ik}^{(r)}\log\mathcal L_{ik,n}({\bm \vartheta})\\
&=\mathcal Q_1({\bm \omega})+\mathcal Q_2({\bm \vartheta})\, .
\end{aligned}
\end{equation}
In the above, 
$\widetilde{\chi}_{ik}^{(r)}
=\mathbb E\left\{\chi_{ik}\mid O_i,\widehat{\bm \theta}^{(r)}\right\}
=\mathbb P\left\{G_i=k\mid O_i,\widehat{\bm \theta}^{(r)}\right\}\, ,
$
and by the Bayes' rule, one can show that 
$$
\widetilde{\chi}_{ik}^{(r)} = \frac{\widehat{\omega}_k^{(r)}
\mathcal L_{ik,n}\{\widehat{\bm \vartheta}^{(r)}\}} {\sum_{l=1}^K\widehat{\omega}_l^{(r)}
\mathcal L_{il,n}\{\widehat{\bm \vartheta}^{(r)}\}}\, .
$$ 

In the M-step, we maximize \eqref{Qfunc} with respect to ${\bm \theta}$ to obtain $\widehat{\bm \theta}^{(r+1)}$. Since ${\bm \omega}$ 
appears only in $\mathcal Q_1$ and ${\bm \vartheta}$ appears only in $\mathcal Q_2$, they can be updated separately. 
The maximization of $\mathcal Q_1$ subject to $\sum_{k=1}^K\omega_k=1$ gives
\begin{equation}\label{updateomega}
\widehat{\omega}_k^{(r+1)} = \frac{1}{n}\sum_{i=1}^n\widetilde{\chi}_{ik}^{(r)}\, ,
\qquad k=1,\ldots,K\, .
\end{equation} 
The remaining parameters can be updated jointly by
$$
\widehat{\bm \vartheta}^{(r+1)} = \arg\max_{\bm \vartheta} \sum_{i=1}^n\sum_{k=1}^K \widetilde{\chi}_{ik}^{(r)} \log\mathcal L_{ik,n}({\bm \vartheta})\, .
$$ 
Alternatively, to improve numerical stability, we employ a blockwise maximization procedure in the numerical studies below. Specifically, one can perform the following
\begin{equation}\label{updatebeta}
\widehat{\bm \beta}^{(r+1)} = \arg\max_{\bm \beta} \sum_{i=1}^n\sum_{k=1}^K \widetilde{\chi}_{ik}^{(r)} \log\mathcal L_{ik,n}\left\{ {\bm \beta}, \, \widehat{\bm \zeta}^{(r)}, \, \widehat{\bm \psi}^{(r)} \right\}\, ,
\end{equation}
\begin{equation}\label{updatezeta}
\widehat{\bm \zeta}^{(r+1)} = \arg\max_{\bm \zeta} \sum_{i=1}^n\sum_{k=1}^K \widetilde{\chi}_{ik}^{(r)} \log\mathcal L_{ik,n}\left\{ \widehat{\bm \beta}^{(r+1)}, \, {\bm \zeta}, \, \widehat{\bm \psi}^{(r)} \right\}\, ,
\end{equation}
and
\begin{equation}\label{updatepsi}
\widehat{\bm \psi}^{(r+1)} = \arg\max_{\bm \psi} \sum_{i=1}^n\sum_{k=1}^K \widetilde{\chi}_{ik}^{(r)} \log\mathcal L_{ik,n}\left\{ \widehat{\bm \beta}^{(r+1)}, \, \widehat{\bm \zeta}^{(r+1)},\, {\bm \psi} \right\}\, ,
\end{equation}
where ${\bm \psi} = ({\bm \psi}_1^\top, \ldots, {\bm \psi}_K^\top)^\top$ with ${\bm \psi}_k = (\psi_{k0}, \ldots, \psi_{km})^\top$.

In summary, the EM algorithm iteratively updates \eqref{updateomega}--\eqref{updatepsi} until convergence.  For the convergence,
one may be based on either the change in the parameter estimates or the change in the observed-data log likelihood function 
being smaller than a prespecified value.  In the numerical studies below, we use the latter criterion with a tolerance of $10^{-5}$. 
It should be noted that the EM algorithm may converge to a local maximum and to deal with this, 
we suggest trying multiple initial values and comparing the observed-data log likelihood function values. 
After obtaining the final estimator $\widehat{\bm \theta}$, the functional coefficient for subgroup $k$ can be estimated by $\widehat{\gamma}_k(s)=\widehat{\bm \phi}_k^\top{\bm b}(s)$ for $k=1,\ldots,K$.

In the above, we have assumed that $K$ is known and this could be the case if there exists subject-matter knowledge or existing literature. 
It is apparent that this may not be true in practice and in this latter case, the following data-driven procedure for selecting $K$ can be used. 
Specifically, suppose that there exists a fixed finite set of candidate values $\mathcal K\subset\mathbb N$.  For each $K\in\mathcal K$,
we apply the proposed estimation procedure above to obtain $\widehat{\bm \theta}_{[K]}$.  Then one can estimate the number of subgroups by
$\widehat K=\arg\min_{K\in\mathcal K}\mathrm{BIC}(K)$, 
where
\begin{equation}\label{BIC}
\mathrm{BIC}(K)=-2\ell_n(\widehat{\bm \theta}_{[K]})+d_K\log n
\end{equation}
with $d_K=p_1+K(p_2+q_n)+K(m+1)+(K-1)$, the total number of free parameters. Alternatively, one may apply 
the AIC or cross-validation criterion.

\section{Asymptotic Properties}

Now we establish the asymptotic properties of the proposed estimator $\widehat{\bm \theta}$. 
Let ${\bm \theta}_0$ denote the true value of ${\bm \theta}$ and $\|{\bm a}\|$ the  Euclidean norm
of a finite-dimensional vector ${\bm a}$. For $\gamma\in L_2[0,1]$, define
$$
\|\gamma\|_{\mathcal{H}}^2=\mathbb E\left\{\int_0^1X(s)\gamma(s)\,ds\right\}^2 =\int_0^1\int_0^1\gamma(s)C_X(s,t)\gamma(t)\,ds\,dt\, ,
$$
where $C_X(s,t)=\operatorname{cov}\{X(s),X(t)\}$. For a function $\Lambda_k$, define
$$
\|\Lambda_k\|_{\Lambda}^2=\int\{\Lambda_k(l)^2+\Lambda_k(r)^2\}\,dF(l,r)\, ,
$$
where $F(l,r)$ denotes the joint distribution of $(L,R)$. 
Furthermore, for two vectors ${\bm \theta}$ and $\widetilde{\bm \theta}$, define
$$
d^2({\bm \theta},\widetilde{\bm \theta}) = \|{\bm \beta}-\widetilde{\bm \beta}\|^2 +\|{\bm \omega}-\widetilde{\bm \omega}\|^2
+\sum_{k=1}^K\left\{ \|{\bm \alpha}_k-\widetilde{\bm \alpha}_k\|^2 +\|\gamma_k-\widetilde{\gamma}_k\|_\mathcal{H}^2
+\|\Lambda_k-\widetilde{\Lambda}_k\|_{\Lambda}^2 \right\}\, .
$$
It is worth noting that the latent subgroup labels are identifiable only up to a permutation, and the value of this distance may change under 
a relabeling even though the fitted mixture model remains unchanged. We therefore assume that the estimated subgroup labels have been aligned with the true labels. 
Let $K_0$ denote the true number of subgroups.  We first establish the consistency and asymptotic normality of the proposed estimator 
assuming that $K=K_0$ is known. 

\begin{theorem}\label{theorem1}
Suppose that Conditions~C1--C7 given in the Web Appendix A of the Supplementary Material hold. Then we have that 
$d(\widehat{\bm \theta},{\bm \theta}_0)\overset{p}{\longrightarrow}0$ as $n\to\infty$. 
\end{theorem}

\begin{theorem}\label{theorem2}
Suppose that Conditions~C1--C7 given in the Web Appendix A of the Supplementary Material hold.  Also suppose that
 $q_n\asymp n^\rho$, $m\asymp n^\nu$, $\max(\rho,\nu)<1/2$, $\rho r_\gamma>1/4$, and $\nu r_\Lambda>1/2$, 
 where $r_\gamma$ and $r_\Lambda$ are defined in Condition~C3.  Then we have that 
 $\sqrt n(\widehat{\bm\wp}-{\bm\wp}_0)$ converges in distribution to a zero-mean multivariate normal distribution as $n\to\infty$.
\end{theorem}

The proofs of the results above are sketched in the Web Appendix B of the Supplementary Material.  
In the next theorem, we consider the situation where $K_0$ is not known but there exists a finite candidate set $\mathcal K$ 
containing $K_0$ and establish the consistency of the resulting estimator with respect to the oracle estimator. 

\begin{theorem}\label{theorem3}
Suppose that Conditions~C1--C7 given in the Web Appendix A of the Supplementary Material hold and that $K_0\in\mathcal K$. Then we have that
(i) $\mathbb{P}(\widehat K<K_0)\rightarrow 0$;
(ii) $\|\widehat{\bm\beta}_{[\widehat K]}-\widehat{\bm\beta}\|\overset{p}{\longrightarrow}0$. Moreover, for each $k\in\{1,\ldots,K_0\}$, there exists an index $j_n(k)\in\{1,\ldots,\widehat K\}$ such that $\|\widehat{\bm\alpha}_{[\widehat K],j_n(k)}-\widehat{\bm\alpha}_k\|\overset{p}{\longrightarrow}0$, where $\widehat{\bm\beta}$ and $\widehat{\bm\alpha}=(\widehat{\bm\alpha}_1^\top,\ldots,\widehat{\bm\alpha}_{K_0}^\top)^\top$ denote the oracle estimators discussed in Theorem~\ref{theorem1}.
\end{theorem}

The proof of the results above is sketched in the Web Appendix B of the Supplementary Material. 
Note that Theorem~\ref{theorem3} suggests that the selected model will not asymptotically underestimate the true number of subgroups. 
Moreover, for each true subgroup, there exists at least one fitted component with $\widehat K$ whose regression parameter 
can be matched to the corresponding oracle component. In other words, the selected model may contain additional fitted components, which may reflect over-splitting of the underlying subgroup structure.

It is apparent that to make inference about ${\bm \wp}$, or ${\bm \beta}$ and ${\bm \alpha}$, one needs to estimate their covariance matrices.
For this, since the closed-forms of the covariance matrices are analytically cumbersome, we suggest to use the nonparametric bootstrap procedure. Specifically, we draw a bootstrap sample from the original data and 
apply the proposed estimation procedure to obtain a bootstrap estimate $\widehat{\bm \theta}^{*}$. Repeating this procedure $B$ times yields the bootstrap estimates $\{\widehat{\bm \theta}^{*(1)},\ldots,\widehat{\bm \theta}^{*(B)}\}$. The standard error of each parameter of interest can then be estimated by the sample standard deviation of its corresponding bootstrap estimates. The pointwise standard errors for the subgroup-specific functional coefficients can also be obtained similarly from $\{\widehat{\gamma}_k^{*(1)}(s),\ldots,\widehat{\gamma}_k^{*(B)}(s)\}$ for each $s\in[0,1]$ and $k=1,\ldots, K$.

\section{A Simulation Study}

In this section, we present some results obtained from an extensive simulation study conducted to evaluate the empirical performance 
of the estimation procedure proposed in the previous sections. In the study, we considered the two-dimensional scalar covariate 
$\bm{W}_i = (W_{i1}, W_{i2})^\top$ generated from the multivariate standard normal distribution and the one-dimensional scalar covariate $Z_i$ 
following the standard normal distribution. For the functional coefficient $X_i(\cdot)$, we used a setup similar to that of \citet{qu2016optimal} 
and \citet{liu2025efficient}. Specifically, we generated $X_i(s) = \sum_{j=1}^{50} \xi_j U_{ij} \varphi_j(s)$, where the $U_{ij}$'s 
were independently sampled from $\mathrm{Uniform}(0, 1)$, $\xi_j = (-1)^{j+1} j^{-1/2}$, $\varphi_1(s) \equiv 1$, and $\varphi_{j+1}(s) = \sqrt{2} \cos(j\pi s)$ for $j \ge 1$.

For the subgroup structure, we considered four scenarios. In scenario (A), we considered a homogeneous or single-subgroup situation 
with $\alpha_1 = 1$ and $\Lambda_1(t) = t$. The functional coefficient was set to $\gamma_1(s) = \sum_{j=1}^{50} (-1)^j j^{-3/2} \varphi_j(s)$. 
In scenario (B), we assumed two balanced subgroups with $\mathbb{P}(G_i = 1) = \mathbb{P}(G_i = 2) = 0.5$, and the heterogeneous coefficients were set to $\bm{\alpha} = (1, -1)^\top$. In subgroup I, we set $\Lambda_1(t) = t$ and $\gamma_1(s) = \sum_{j=1}^{50} (-1)^j j^{-3/2} \varphi_j(s)$,
while in subgroup II, we set $\Lambda_2(t) = t^2$ and $\gamma_2(s) = -\sum_{j=1}^{50} (-1)^j j^{-3/2} \varphi_j(s)$. 
In scenario (C), we mimic scenario (B) but assumed that both subgroups share the same functional coefficient with $\gamma_1(s) = \gamma_2(s) = \sum_{j=1}^{50} (-1)^j j^{-3/2} \varphi_j(s)$. This allows us to assess whether the proposed method can still distinguish subgroups when the functional coefficients are not subgroup-specific. 
In scenario (D), we again mimic scenario (B) but considered an unbalanced setting with $\mathbb{P}(G_i = 1) = 0.4$ and $\mathbb{P}(G_i = 2) = 0.6$. 
Given the scenarios above, the failure times $T_i$'s were generated from model \eqref{haz} with the true regression coefficient $\bm{\beta} = (1, -1)^\top$.

To generate the censoring intervals $(L_i, R_i]$'s, following \citet{zhou2017sieve}, we mimicked a medical follow-up study in which each subject was scheduled to be examined at $\iota$ equally spaced time points over $(0,\tau)$. At each scheduled time, the subject was examined for the occurrence of the failure event with probability $0.5$, independently across examination times. For subject $i$, $L_i$ was defined as the last actual examination time before $T_i$, and $R_i$ 
as the first actual examination time after $T_i$. We set $\tau=2.5$ and $\iota=20$, which resulted in right censoring rates approximately  $20\%$ -- $30\%$ across different scenarios. The results presented below are based on sample sizes $n = 400$ or $800$, $m = 3$, $q_n = \max(\lfloor n^{1/4} \rfloor, \, 4)$ and 50 bootstrap samples with 500 replications.

To assess the performance of the proposed method, we considered several criteria. For estimation of the number of subgroups, 
we calculated the mean, median and standard deviation (SD) of $\widehat K$ across simulation replications. For estimation of the regression parameters and functional coefficients, we calculated the root mean squared error between the fitted and true latent linear predictors as
$$
\mathrm{RMSE} = \left[ \frac{1}{n}\sum_{i=1}^n \left\{ \widehat{\eta}_{i,\widehat G_i} - \eta_{0i,G_i} \right\}^2 \right]^{1/2}\, .
$$
In the above, $\widehat{\eta}_{i,\widehat G_i} = {\bm W}_i^\top\widehat{\bm \beta} + {\bm Z}_i^\top\widehat{\bm \alpha}_{\widehat G_i} + \int_{\mathcal I_0}X_i(s)\widehat{\gamma}_{\widehat G_i}(s)\,ds$ with $\widehat G_i=\arg\max_{1\leq k\leq\widehat K}\widetilde{\chi}_{ik}$ and 
$\eta_{0i,G_i} = {\bm W}_i^\top{\bm \beta}_0 + {\bm Z}_i^\top{\bm \alpha}_{0,G_i} + \int_{\mathcal I_0}X_i(s)\gamma_{0,G_i}(s)\,ds$
with $G_i$ denoting the true subgroup label of subject $i$.
Note that this RMSE accounts jointly for estimation errors in the scalar coefficients, functional coefficients, and latent subgroup memberships. 
To further evaluate parameter estimation, we focused on the simulation replications in which the number of subgroups was correctly selected. After aligning the estimated subgroup labels with the true labels, we assessed the recovery of the functional coefficient in subgroup $k$ using
$$
\mathrm{ISE}_k = \left[ \int_0^1 \left\{ \widehat{\gamma}_k(s)-\gamma_{0k}(s) \right\}^2\,ds \right]^{1/2}\, .
$$
For the scalar coefficients, we evaluated their estimation and inference performance using the empirical bias (Bias), average estimated standard error (SD), empirical standard error (SEE), and empirical 95\% coverage probability (CP).

\centerline{
[Tables \ref{tableA1} -- \ref{tableA2} about here.]}

\begin{figure}
     \centering
      \begin{subfigure}[b]{0.48\textwidth}
         \centering
         \includegraphics[width=\textwidth]{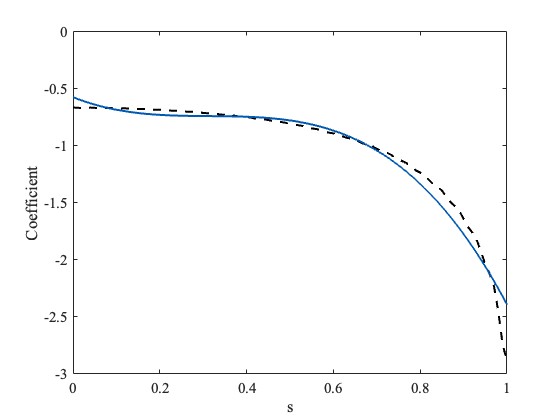}
         \caption{\scriptsize (A): $n = 400$, Subgroup I}
     \end{subfigure}
     \begin{subfigure}[b]{0.48\textwidth}
         \centering
         \includegraphics[width=\textwidth]{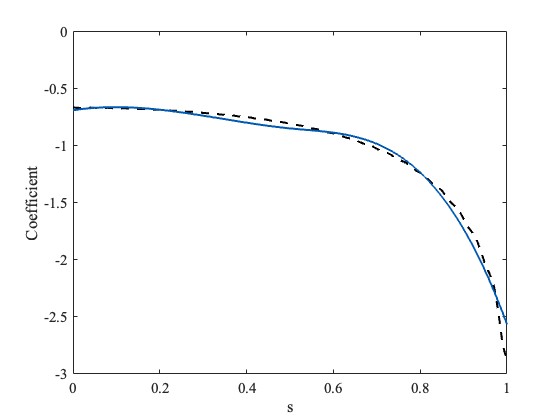}
         \caption{\scriptsize (A): $n = 800$, Subgroup I}
     \end{subfigure}
      \begin{subfigure}[b]{0.48\textwidth}
         \centering
         \includegraphics[width=\textwidth]{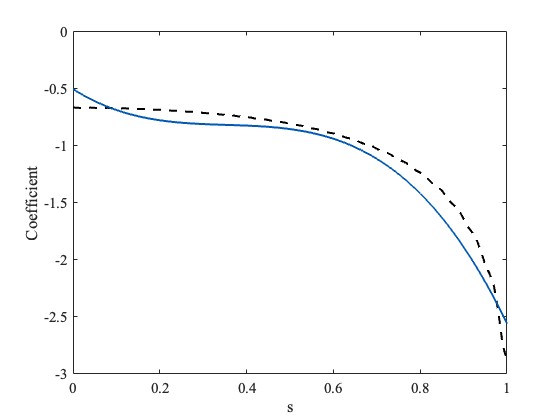}
         \caption{\scriptsize (B): $n = 400$, Subgroup I}
     \end{subfigure}
     \begin{subfigure}[b]{0.48\textwidth}
         \centering
         \includegraphics[width=\textwidth]{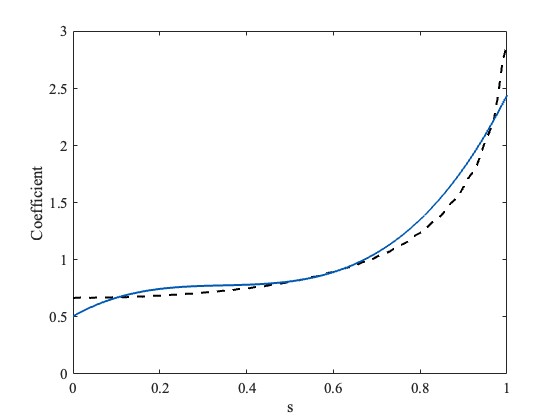}
         \caption{\scriptsize (B): $n = 400$, Subgroup II}
     \end{subfigure}
      \begin{subfigure}[b]{0.48\textwidth}
         \centering
         \includegraphics[width=\textwidth]{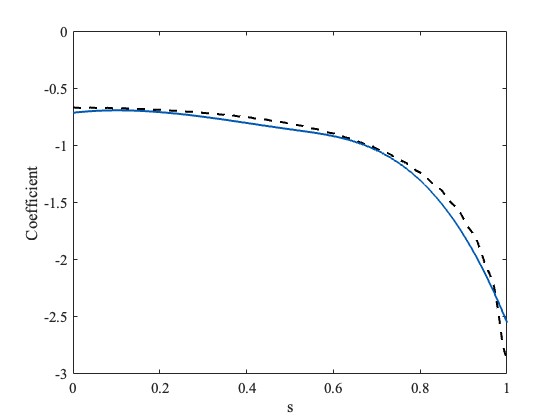}
         \caption{\scriptsize (B): $n = 800$, Subgroup I}
     \end{subfigure}
     \begin{subfigure}[b]{0.48\textwidth}
         \centering
         \includegraphics[width=\textwidth]{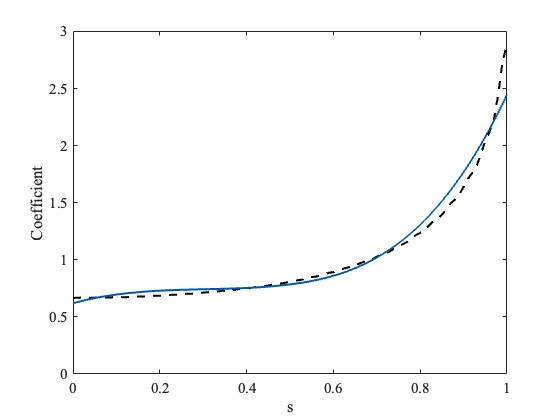}
         \caption{\scriptsize (B): $n = 800$, Subgroup II}
     \end{subfigure}
	\caption{Estimated coefficient function $\widehat{\gamma}(s)$ under Scenarios (A) and (B): true curve (dashed), estimated curve (solid)}
	\label{figureA}
\end{figure}

\centerline{
[Figure \ref{figureA} about here.]}

\begin{figure}
     \centering
      \begin{subfigure}[b]{0.48\textwidth}
         \centering
         \includegraphics[width=\textwidth]{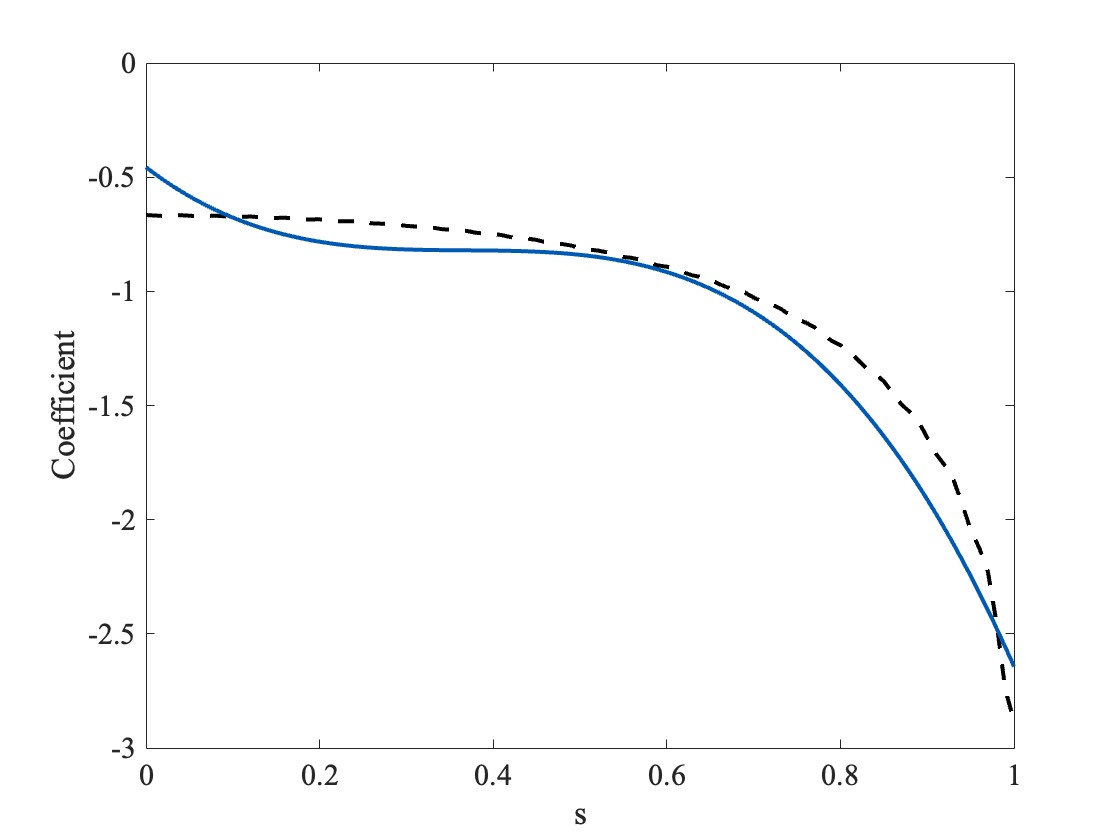}
         \caption{\scriptsize (C): $n = 400$, Subgroup I}
     \end{subfigure}
     \begin{subfigure}[b]{0.48\textwidth}
         \centering
         \includegraphics[width=\textwidth]{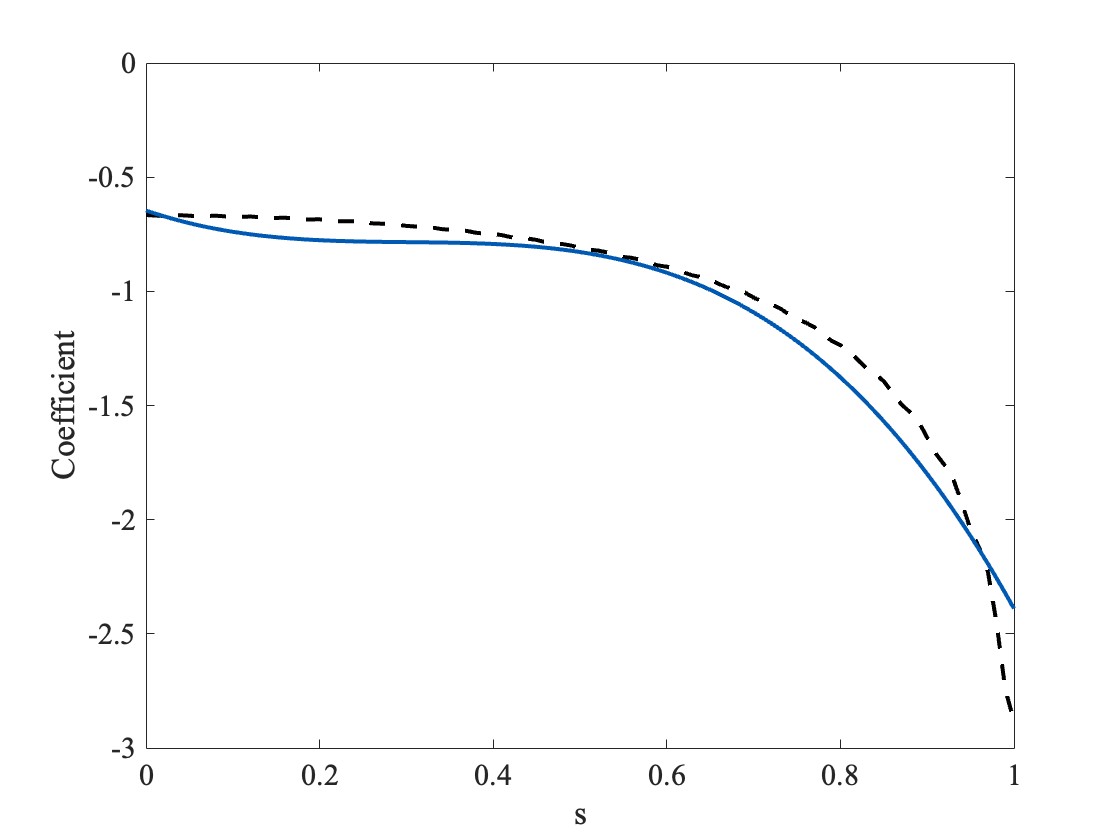}
         \caption{\scriptsize (C): $n = 400$, Subgroup II}
     \end{subfigure}
      \begin{subfigure}[b]{0.48\textwidth}
         \centering
         \includegraphics[width=\textwidth]{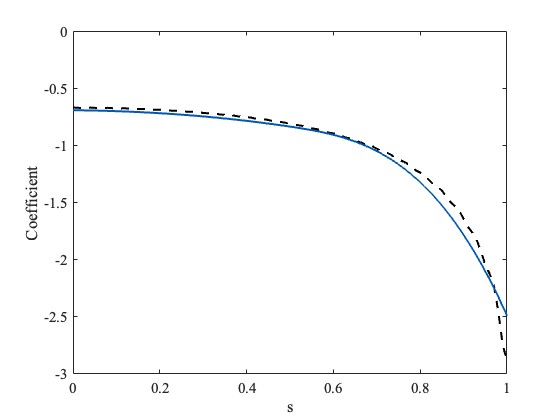}
         \caption{\scriptsize (C): $n = 800$, Subgroup I}
     \end{subfigure}
     \begin{subfigure}[b]{0.48\textwidth}
         \centering
         \includegraphics[width=\textwidth]{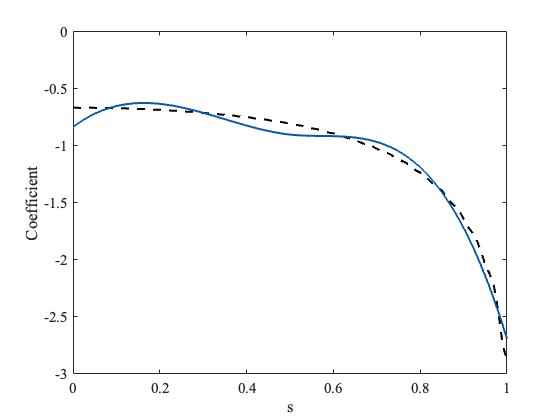}
         \caption{\scriptsize (C): $n = 800$, Subgroup II}
     \end{subfigure}
      \begin{subfigure}[b]{0.48\textwidth}
         \centering
         \includegraphics[width=\textwidth]{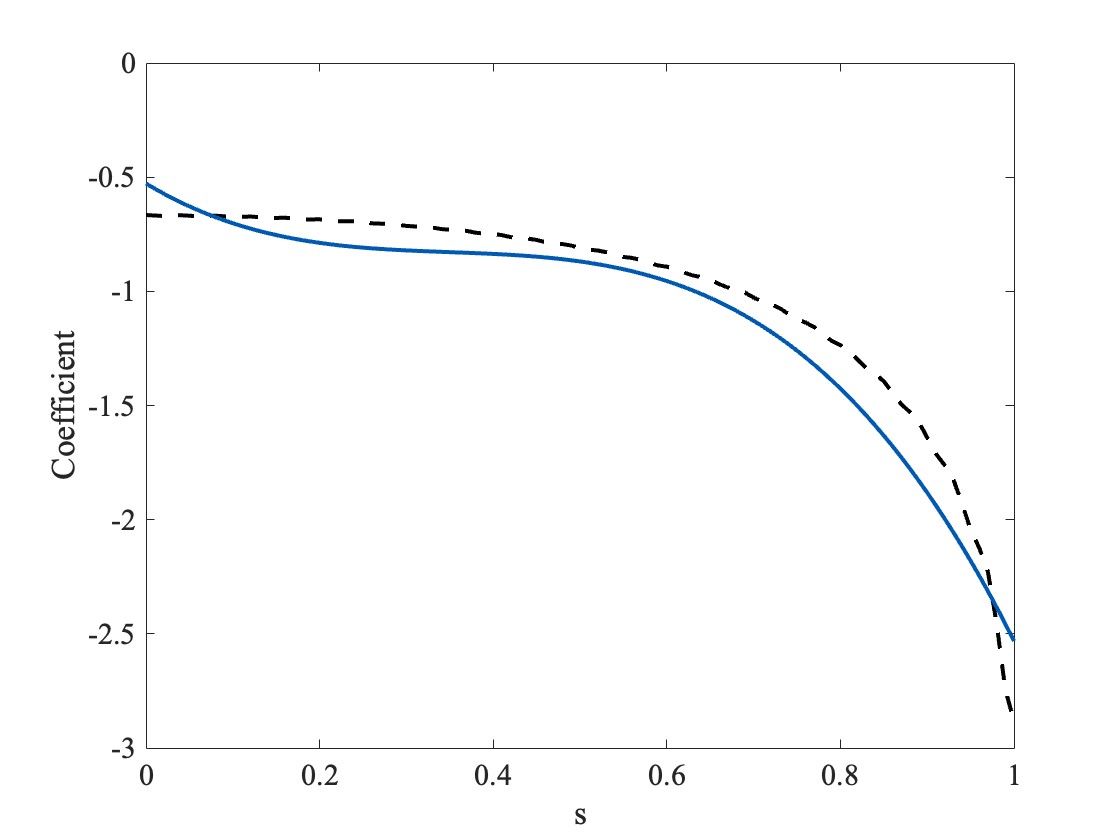}
         \caption{\scriptsize (D): $n = 400$, Subgroup I}
     \end{subfigure}
     \begin{subfigure}[b]{0.48\textwidth}
         \centering
         \includegraphics[width=\textwidth]{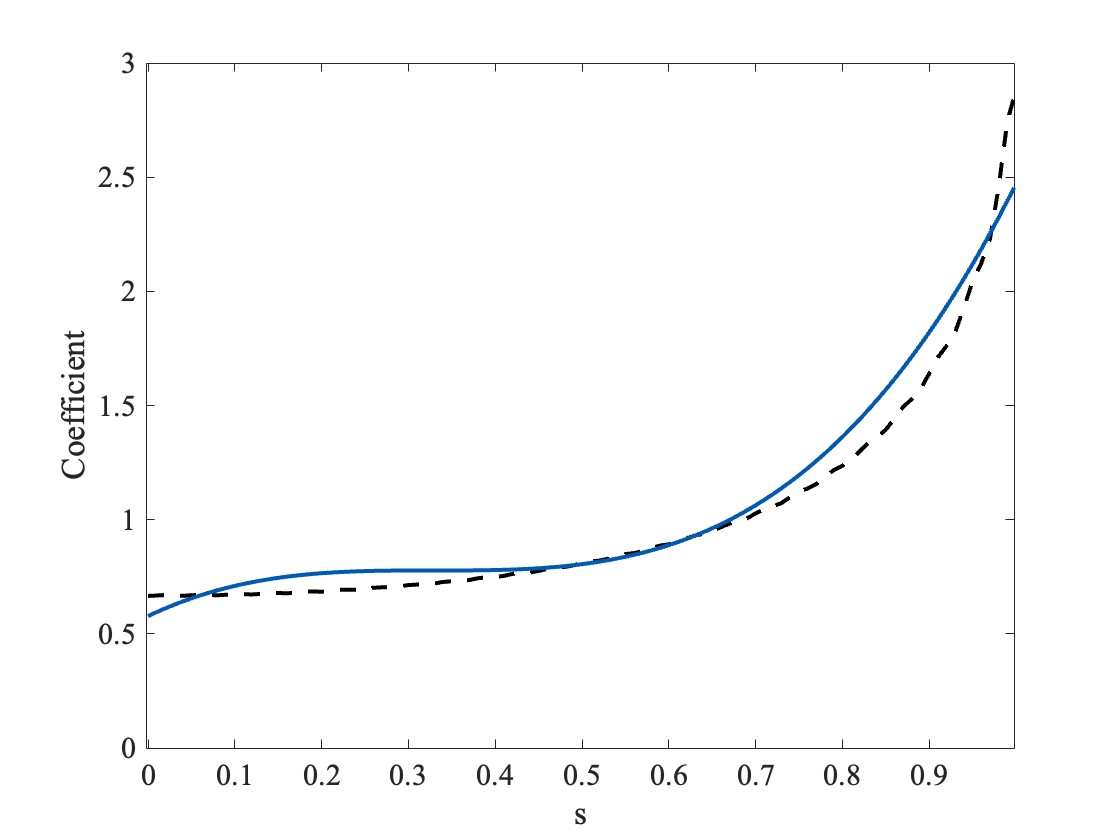}
         \caption{\scriptsize (D): $n = 400$, Subgroup II}
     \end{subfigure}
      \begin{subfigure}[b]{0.48\textwidth}
         \centering
         \includegraphics[width=\textwidth]{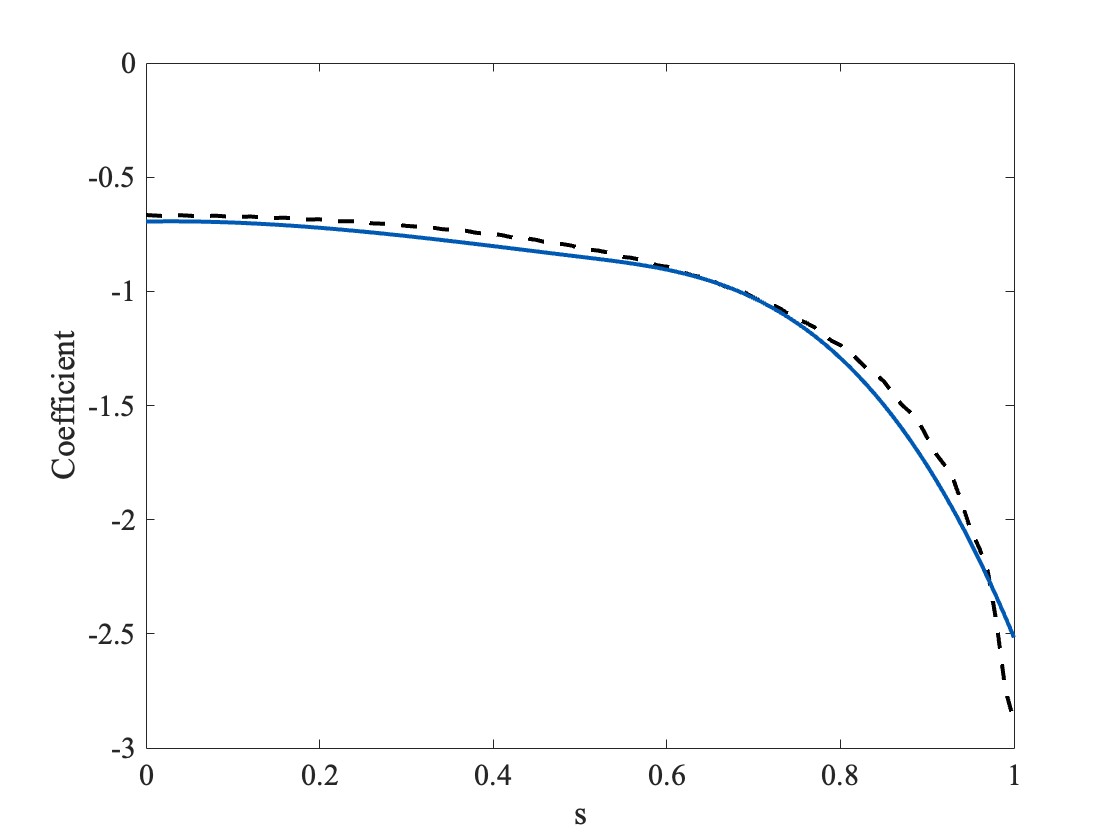}
         \caption{\scriptsize (D): $n = 800$, Subgroup I}
     \end{subfigure}
     \begin{subfigure}[b]{0.48\textwidth}
         \centering
         \includegraphics[width=\textwidth]{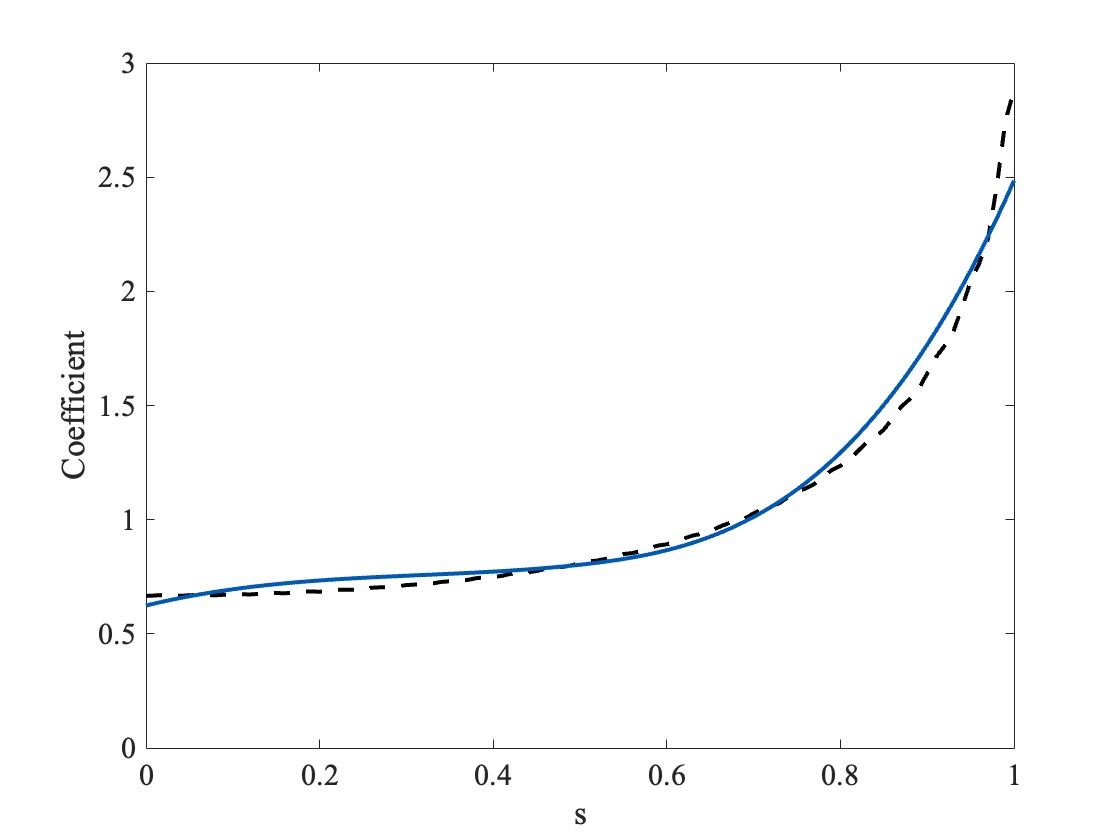}
         \caption{\scriptsize (D): $n = 800$, Subgroup II}
     \end{subfigure}
	\caption{Estimated coefficient function $\widehat{\gamma}(s)$ under Scenarios (C) and (D): true curve (dashed), estimated curve (solid)}
	\label{figureC}
\end{figure}

\centerline{
[Figure \ref{figureC} about here.]}

Table~\ref{tableA1} presents the results on the estimation of the number of subgroups along with several other performance measures 
described above under four scenarios, while Table~\ref{tableA2} gives the results on the estimation of scalar coefficients.  
One can see that the estimated number of subgroups is close to the true value and the probability of correctly selecting the number of 
subgroups increased with the sample size.  Moreover, the scalar coefficient estimates exhibited little bias and the estimated standard errors 
are close to the corresponding empirical standard errors.  The empirical coverage probabilities are also close to the nominal $95\%$ level, 
indicating that the normal approximation seems to be reasonable.  Figures~\ref{figureA} -- \ref{figureC} provide the estimated functional 
coefficient $\widehat{\gamma}(s)$ for $n=400$ and $800$, and one can see that the average estimated curves closely follow the corresponding true curves.

Compared with the balanced setting in scenario~(B), we can see that the unbalanced mixing proportions in scenario~(D) had little effect on subgroup selection or parameter estimation. Under scenario~(C), however, the subgroup selection and estimation are slightly less accurate than under scenario~(B). This is expected because the two subgroups in 
scenario~(C) are less distinguishable where they share the same functional coefficients. The results also suggest that allowing subgroup-specific functional coefficients when the true functional effect is common may lead to a slight loss of estimation accuracy. Nevertheless, this loss diminishes as the sample size increases, and the proposed method continues to perform well.

\section{An Application}

Now we apply the estimation procedure proposed in the preceding sections to the Alzheimer's Disease Neuroimaging Initiative (ADNI) discussed above. 
In the study, the participants were classified at baseline into three cognitive status groups, cognitively normal, mild cognitive impairment (MCI), and 
Alzheimer's disease (AD).   As mentioned before, among others, one variable of interest is the time from the baseline visit to conversion to AD
and for it, only interval-censored data are available due to the nature of the study \citep{li2020penalized, lou2026semiparametric}. 
For the analysis, following \citet{li2020penalized} and \citet{lou2026semiparametric}, we will focus on the 372 participants in the MCI group 
and are interested in the existence of subgroups with different rates of disease progression or different covariate effects on the risk of developing AD. 

More specifically, we will focus on the baseline age and two neuropsychological test measures, RAVLTi and LDELt, 
as scalar covariates. RAVLTi denotes the Rey Auditory Verbal Learning Test immediate recall score, which measures initial learning capacity, 
short-term auditory memory, and verbal information encoding, while LDELt denotes the Logical Memory Delayed Recall Total Score, 
which measures delayed recall of a previously presented story or logical memory passage. The MCI diagnosis 
was defined in part using education-adjusted performance on the WMS-R Logical Memory II delayed paragraph recall test \citep{petersen2010alzheimer}, 
which is closely related to the measure represented by LDELt. Therefore, the variation in LDELt across the subjects may be restricted at baseline, 
which can limit its marginal ability to predict subsequent AD conversion. At the same time, previous studies have reported different findings concerning the independent predictive role of delayed logical memory.  For example, \cite{silva2012comparison} found that logical memory delayed recall was not significant in a multivariable Cox model combining several memory measures, whereas \cite{gomar2011utility} retained logical memory delayed recall in their final prediction model. \cite{gomar2014extension} further reported that delayed recall was significant in a cognitive marker model but was not retained in the final combined model. These different findings suggest that the additional prognostic information carried by delayed recall may differ across patient groups and thus motivate us to examine whether its effect varies across latent subgroups. 

For the functional covariate, we consider hippocampal shape, for which radial distance measurements were extracted from approximately 30,000 surface points across both hippocampi \citep{kong2018flcrm, lou2026semiparametric}. These radial distances, defined from each surface vertex to the medial axis, provide information on both hippocampal size and local shape. Previous medical studies have suggested that there may be significant subgroup differences in the association between hippocampal radial distance and disease progression or cognitive decline \citep{apostolova2006conversion,carmichael2012localized}.

\centerline{
[Table \ref{tableappADNI} about here.]}

\begin{figure}
	\centering
	\includegraphics[width=0.9\textwidth]{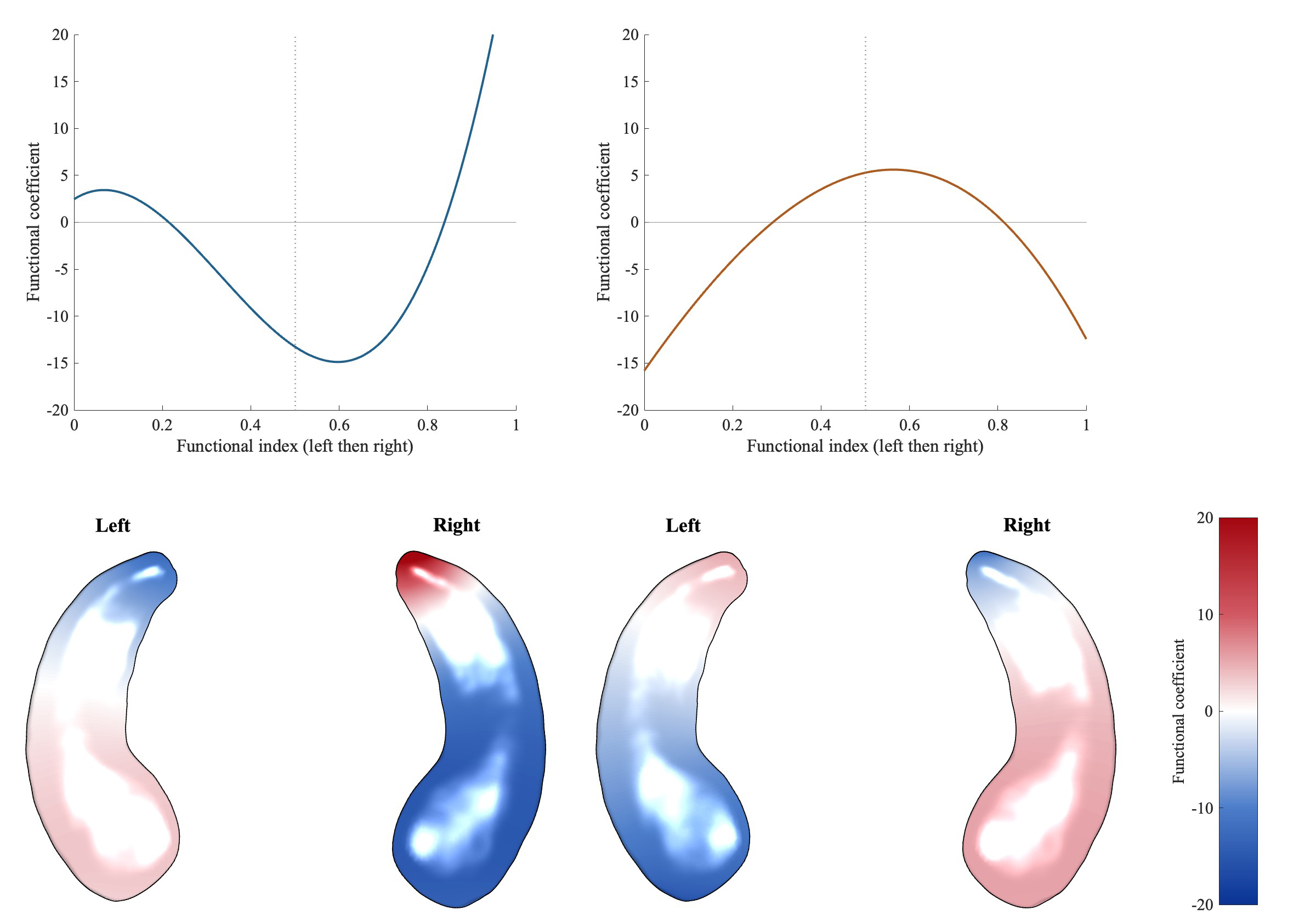}
	\caption{Estimated coefficient functions and their projections onto the hippocampus (left panel: Subgroup 1; right panel: Subgroup 2)}
	\label{fighappKequal2}
\end{figure}

\centerline{
[Figure \ref{fighappKequal2} about here.]}

To apply the proposed method, based on the discussion above,
we include AGE and RAVLTi in ${\bm W}$ and LDELt in ${\bm Z}$ and set $m=3$ and $q_n=4$, consistent with the choice $q_n=\max\{\lfloor n^{1/4}\rfloor,4\}$ 
used in the simulation study. We considered the candidate set $\mathcal K=\{1,2,3,4,5\}$ and the BIC yielded $\hat{K} = 2$ with estimated subgroup weights $\widehat{\bm\omega}=(\widehat{\omega}_1,\widehat{\omega}_2)^\top=(0.4935,0.5065)^\top$. That is, the proposed method identified 
two latent subgroups.  Note that a related mixture survival analysis by \cite{feng2025latent}, which allowed subgroup-specific imaging effects, also identified two latent classes. 
Table~\ref{tableappADNI} presents the estimation results for the baseline factors, including the estimated effects (EST), standard errors (SE), and $p$-values for testing whether each effect is zero (PVAL). They
suggest that RAVLTi had a significant negative common effect, whereas AGE was not significant. 
The negative association of RAVLTi is consistent with the finding reported by \cite{zhang2025post}.

Furthermore, LDELt had a significant negative effect in subgroup~2 but was not significant in subgroup~1.  This subgroup-specific result provides one possible explanation for the different findings concerning delayed logical memory in the existing literature, since an effect that is strong in only part of the population may be attenuated when a single common effect is imposed on all subjects. 
The estimated functional coefficients for the two subgroups and their projections onto the hippocampal surfaces are presented
in Figure~\ref{fighappKequal2}, which indicates that the hippocampal effects were clearly different between the two subgroups, with distinct regional patterns across the left and right hippocampi. These differences suggest that the two MCI subgroups may represent distinct hippocampal morphological phenotypes associated with subsequent AD conversion.

\begin{figure}
	\centering
	\includegraphics[width=0.9\textwidth]{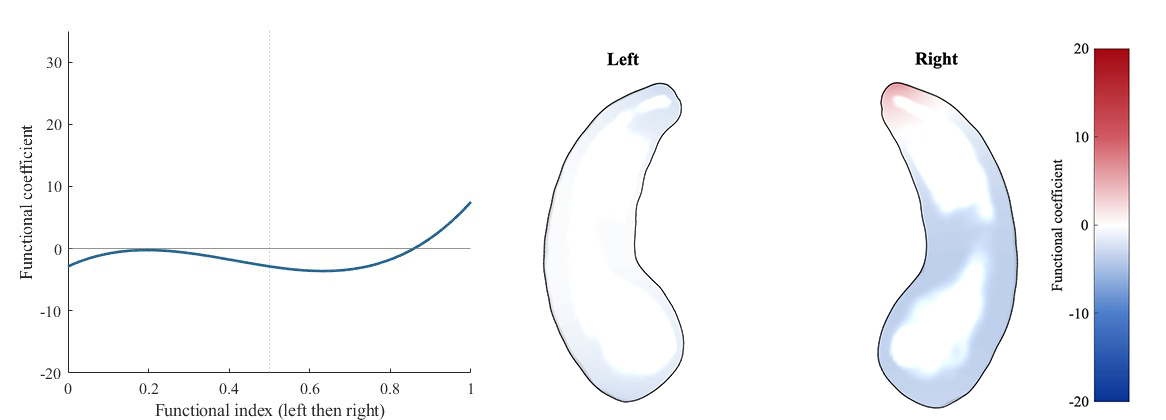}
	\caption{Naive estimated coefficient function and its projection onto the hippocampus}	
	\label{fighappKequal1}
\end{figure}

\centerline{
[Figure \ref{fighappKequal1} about here.]}

For comparison, we also analyzed the data assuming $K = 1$ as the naive analysis without latent subgroup structure and provide 
the corresponding results also in Table~\ref{tableappADNI} along with Figure~\ref{fighappKequal1}.  One can see that the analysis also identified 
protective effects of RAVLTi and LDELt but additionally suggests a significant protective effect of AGE. This would imply that older patients 
have a lower risk of AD conversion, which is not consistent with the established understanding of AD progression. Moreover, the estimated functional 
coefficient is close to zero across most of the hippocampal surface.  One possible reason for this may result from combining the subjects 
from the two latent subgroups into a single population, which can obscure the subgroup-specific effects of both the scalar and functional covariates.

\section{Discussion and Concluding Remarks}

In this paper, we considered estimation of a heterogeneous and functional Cox proportional hazards model 
and proposed a sieve maximum likelihood estimation procedure that can estimate the subgroup structure and covariate effects simultaneously. 
Unlike the existing methods, the proposed approach can simultaneously allow or deal with interval censoring, latent subgroup structure and functional covariates. 
In the method, B-spline basis functions were used to approximate the subgroup-specific functional coefficients and Bernstein polynomials 
to approximate the unknown baseline cumulative hazard functions. The asymptotic properties of the resulting estimators were established, 
and the simulation results suggested that the proposed method worked well in estimating the model parameters and recovering the subgroup structure. 
Also the application to the ADNI study provided some new insights about the possible heterogeneous effects of clinical measurements and hippocampal shape on AD conversion.

Note that for simplicity, we only considered one functional covariate above and it is apparent that more functional covariates may exist. 
The proposed method can be extended directly to the situation by introducing a separate functional coefficient for each functional covariate and applying the same sieve approximation. It should also be noted that Bernstein polynomials were used to approximate the unknown baseline cumulative hazard functions. Other smooth basis functions, including B-splines, could also be considered. One reason for using Bernstein polynomials here is that the required monotonicity and nonnegativity constraints on the baseline cumulative hazard functions can be imposed conveniently through restrictions on the corresponding coefficients.

There also exist several directions for future research.  One is that in the above, we have assumed that the interval censoring is independent or noninformative. In practice, this assumption may not hold when the examination process is related to the failure time of interest \citep{sun2006statistical, sun2022emerging}. For the situation, one may need to jointly model the failure time and the examination process through shared latent variables or copula models. Although the general estimation idea may still be useful, the resulting estimation and computation would be considerably more complicated. Another possible direction is to consider multivariate interval-censored failure time data. The current method focuses on a single failure time, and extending the proposed method to multivariate interval-censored failure outcomes would be of both practical and methodological interest.


\bibliographystyle{apalike}
\bibliography{reference.bib}

\clearpage

\begin{table}
\caption{Simulation results for estimated subgroup information under (A) and (B).}\label{tableA1}
\setlength{\tabcolsep}{1.0mm}{
\fontsize{10}{14}\selectfont
\resizebox{\textwidth}{!}{%
\begin{tabular*}{1.1\textwidth}{@{\extracolsep{\fill}}cccccccccccccccccc @{\extracolsep{\fill}}}
\toprule%
         &     & \multicolumn{3}{c}{Subgroup Number} &  & \multicolumn{3}{c}{RMSE} &  &       & \multicolumn{3}{c}{ISE} &  & \multicolumn{3}{c}{${\bm \omega}$} \\ \cmidrule{3-5} \cmidrule{7-9} \cmidrule{12-14} \cmidrule{16-18} 
Scenario & $n$ & Mean      & Median      & SD        &  & Mean   & Median  & SD    &  & Group & Mean   & Median & SD    &  & Mean      & Median     & SD        \\ \midrule 
(A)      & 400 & 1          & 1          & 0.000     &  & 0.193  & 0.183   & 0.062 &  & I     & 0.356  & 0.339  & 0.129 &  & ---       & ---        & ---       \\ \cmidrule{2-18} 
         & 800 & 1          & 1          & 0.000     &  & 0.144  & 0.138   & 0.044 &  & I     & 0.304  & 0.293  & 0.107 &  & ---       & ---        & ---       \\ \midrule 
(B)      & 400 & 1.970      & 2          & 0.171     &  & 0.879  & 0.856   & 0.121 &  & I     & 0.738  & 0.707  & 0.295 &  & 0.494     & 0.493      & 0.047     \\ \cmidrule{11-18} 
         &     &            &            &           &  &        &         &       &  & II    & 0.668  & 0.650  & 0.258 &  & 0.506     & 0.507      & 0.047     \\ \cmidrule{2-18} 
         & 800 & 2          & 2          & 0.000     &  & 0.813  & 0.812   & 0.061 &  & I     & 0.596  & 0.577  & 0.227 &  & 0.497     & 0.496      & 0.033     \\ \cmidrule{11-18} 
         &     &            &            &           &  &        &         &       &  & II    & 0.547  & 0.518  & 0.207 &  & 0.503     & 0.504      & 0.033    \\ 
\bottomrule
\end{tabular*}
}}
\end{table}

\begin{table}
\caption{Simulation results for the estimated scalar parameters under (A) and (B).}\label{tableA2}
\setlength{\tabcolsep}{1.0mm}{
\fontsize{9}{12}\selectfont
\begin{tabular*}{\textwidth}{@{\extracolsep{\fill}}ccccccccccc @{\extracolsep{\fill}}}
\toprule%
         &            & \multicolumn{4}{c}{$n=400$}       &  & \multicolumn{4}{c}{$n=800$}       \\ \cmidrule{3-6} \cmidrule{8-11} 
Scenario & Parameter  & Bias    & SD     & SEE    & CP    &  & Bias    & SD     & SEE    & CP    \\ \midrule 
(A)      & $\beta_1$  & 0.0167  & 0.0844 & 0.0879 & 0.952 &  & 0.0121  & 0.0641 & 0.0604 & 0.936 \\
         & $\beta_2$  & -0.0180 & 0.0846 & 0.0873 & 0.962 &  & -0.0063 & 0.0615 & 0.0599 & 0.946 \\
         & $\alpha_1$ & 0.0226  & 0.0881 & 0.0876 & 0.936 &  & 0.0066  & 0.0619 & 0.0595 & 0.938 \\ \midrule 
(B)      & $\beta_1$  & 0.0492  & 0.1070 & 0.1267 & 0.961 &  & 0.0212  & 0.0725 & 0.0765 & 0.954 \\
         & $\beta_2$  & -0.0514 & 0.1092 & 0.1262 & 0.961 &  & -0.0197 & 0.0702 & 0.0765 & 0.962 \\
         & $\alpha_1$ & 0.0937  & 0.1856 & 0.2643 & 0.979 &  & 0.0332  & 0.1180 & 0.1437 & 0.972 \\
         & $\alpha_2$ & -0.0313 & 0.1754 & 0.2167 & 0.979 &  & -0.0174 & 0.1067 & 0.1280 & 0.984 \\ 
\bottomrule
\end{tabular*}
}
\end{table}

\clearpage

\begin{table}
\caption{Simulation results for estimated subgroup information under (C) and (D).}\label{tableC1}
\setlength{\tabcolsep}{1.0mm}{
\fontsize{10}{14}\selectfont
\resizebox{\textwidth}{!}{%
\begin{tabular*}{1.1\textwidth}{@{\extracolsep{\fill}}cccccccccccccccccc @{\extracolsep{\fill}}}
\toprule%
         &     & \multicolumn{3}{c}{Subgroup Number} &  & \multicolumn{3}{c}{RMSE} &  &       & \multicolumn{3}{c}{ISE} &  & \multicolumn{3}{c}{${\bm \omega}$} \\ \cmidrule{3-5} \cmidrule{7-9} \cmidrule{12-14} \cmidrule{16-18} 
Scenario & $n$ & Mean      & Median      & SD        &  & Mean   & Median  & SD    &  & Group & Mean   & Median & SD    &  & Mean      & Median     & SD        \\ \midrule
(C)      & 400 & 1.800      & 2          & 0.400     &  & 0.911  & 0.878   & 0.134 &  & I     & 0.764  & 0.729  & 0.338 &  & 0.493     & 0.493      & 0.054     \\ \cmidrule{11-18} 
         &     &            &            &           &  &        &         &       &  & II    & 0.706  & 0.657  & 0.300 &  & 0.507     & 0.507      & 0.054     \\ \cmidrule{2-18} 
         & 800 & 1.998      & 2          & 0.045     &  & 0.791  & 0.787   & 0.064 &  & I     & 0.603  & 0.576  & 0.221 &  & 0.493     & 0.492      & 0.039     \\ \cmidrule{11-18} 
         &     &            &            &           &  &        &         &       &  & II    & 0.553  & 0.542  & 0.193 &  & 0.507     & 0.508      & 0.039     \\ \midrule 
(D)      & 400 & 1.988      & 2          & 0.109     &  & 0.839  & 0.821   & 0.111 &  & I     & 0.885  & 0.829  & 0.381 &  & 0.400     & 0.401      & 0.048     \\ \cmidrule{11-18} 
         &     &            &            &           &  &        &         &       &  & II    & 0.573  & 0.541  & 0.224 &  & 0.600     & 0.599      & 0.049     \\  \cmidrule{2-18} 
         & 800 & 2          & 2          & 0.000     &  & 0.778  & 0.775   & 0.062 &  & I     & 0.703  & 0.679  & 0.278 &  & 0.402     & 0.401      & 0.032     \\ \cmidrule{11-18} 
         &     &            &            &           &  &        &         &       &  & II    & 0.475  & 0.455  & 0.172 &  & 0.598     & 0.599      & 0.032    \\ 
\bottomrule
\end{tabular*}
}}
\end{table}

\begin{table}
\caption{Simulation results for the estimated scalar parameters under (C) and (D).}\label{tableC2}
\setlength{\tabcolsep}{1.0mm}{
\fontsize{9}{12}\selectfont
\begin{tabular*}{\textwidth}{@{\extracolsep{\fill}}ccccccccccc @{\extracolsep{\fill}}}
\toprule%
         &            & \multicolumn{4}{c}{$n=400$}       &  & \multicolumn{4}{c}{$n=800$}       \\ \cmidrule{3-6} \cmidrule{8-11} 
Scenario & Parameter  & Bias    & SD     & SEE    & CP    &  & Bias    & SD     & SEE    & CP    \\ \midrule 
(C)      & $\beta_1$  & 0.0663  & 0.1041 & 0.1270 & 0.973 &  & 0.0217  & 0.0724 & 0.0767 & 0.960 \\
         & $\beta_2$  & -0.0676 & 0.1035 & 0.1269 & 0.950 &  & -0.0178 & 0.0707 & 0.0766 & 0.962 \\
         & $\alpha_1$ & 0.1420  & 0.2149 & 0.2986 & 0.980 &  & 0.0437  & 0.1289 & 0.1594 & 0.980 \\
         & $\alpha_2$ & -0.0571 & 0.1760 & 0.2299 & 0.975 &  & -0.0104 & 0.1171 & 0.1407 & 0.978 \\ \midrule
(D)      & $\beta_1$  & 0.0457  & 0.1053 & 0.1237 & 0.962 &  & 0.0197  & 0.0697 & 0.0750 & 0.966 \\
         & $\beta_2$  & -0.0468 & 0.1054 & 0.1238 & 0.964 &  & -0.0175 & 0.0688 & 0.0745 & 0.962 \\
         & $\alpha_1$ & 0.0786  & 0.2277 & 0.3222 & 0.978 &  & 0.0245  & 0.1347 & 0.1756 & 0.974 \\
         & $\alpha_2$ & -0.0358 & 0.1492 & 0.1869 & 0.972 &  & -0.0198 & 0.0942 & 0.1087 & 0.966 \\ 
\bottomrule
\end{tabular*}
}
\end{table}

\clearpage

\begin{table}
\caption{Analysis results for the ADNI study.\label{tableappADNI}}
\setlength{\tabcolsep}{1.0mm}{
\fontsize{9}{12}\selectfont
\begin{tabular*}{\textwidth}{@{\extracolsep{\fill}}cccccccc @{\extracolsep{\fill}}}
\toprule%
           & \multicolumn{3}{c}{Naive Method} &  & \multicolumn{3}{c}{Proposed Method} \\ \cmidrule{2-4} \cmidrule{6-8} 
Factor     & EST        & SE       & PVAL     &  & EST         & SE        & PVAL      \\ \midrule 
AGE        & -0.2074    & 0.0882   & 0.0188   &  & -0.3061     & 0.1605    & 0.0565    \\
RAVLTi     & -0.5011    & 0.1029   & 0.0000   &  & -0.7423     & 0.1662    & 0.0000    \\
LDELt (I)  & -0.2831    & 0.0783   & 0.0003   &  & -0.4280     & 0.3307    & 0.1955    \\
LDELt (II) & ---        & ---      & ---      &  & -0.6024     & 0.2924    & 0.0394   \\ 
\bottomrule
\end{tabular*}
}
\end{table}

\end{document}